\documentclass[12pt]{article}
\RequirePackage[OT1]{fontenc}
\RequirePackage{amsthm,amsmath}
\RequirePackage[authoryear]{natbib}

\usepackage{amsmath}
\usepackage{graphicx,psfrag,epsf}
\usepackage{enumerate}
\usepackage{natbib}
\usepackage{multirow}

\usepackage{mathrsfs}
\usepackage{amssymb,amsfonts}
\newtheorem{corollary}{Corollary}
\newtheorem{example}{Example}
\newtheorem{lemma}{Lemma}

\newtheorem{theorem}{Theorem}
\usepackage{xcolor}
\usepackage{epstopdf}
\usepackage[caption=false]{subfig}
\usepackage[latin1]{inputenc}
\usepackage{color}
\usepackage{xcolor}
\usepackage{lineno,hyperref}
\usepackage{multicol}
\usepackage{algorithm}
\usepackage{algpseudocode}
\usepackage{lipsum}
\usepackage{appendix}
\usepackage{amsmath}
\usepackage{graphicx}
\usepackage{lineno}
\usepackage{array}
\usepackage{longtable}
\usepackage{natbib}
\usepackage{float}
\usepackage{comment}
\usepackage{url}
\usepackage{booktabs}
\usepackage{multirow}
\usepackage{natbib}

\begin{document}

\title{\large\bf A Trigamma-free Approach for Computing Information Matrices Related to Trigamma Function}
\author{Zhou Yu$^1$,  Niloufar Dousti Mousavi$^2$, and Jie Yang$^1$\\
		$^1$University of Illinois at Chicago and $^2$University of Chicago}
	
\maketitle

\begin{abstract}
Negative binomial related distributions have been widely used in practice. The calculation of the corresponding Fisher information matrices involves the expectation of trigamma function values which can only be calculated numerically and approximately. 
In this paper, we propose a trigamma-free approach to approximate the expectations involving the trigamma function, along with theoretical upper bounds for approximation errors. We show by numerical studies that our approach is highly efficient and much more accurate than previous methods.  
We also apply our approach to compute the Fisher information matrices of zero-inflated negative binomial (ZINB) and beta negative binomial (ZIBNB) probabilistic models, as well as ZIBNB regression models.
\end{abstract}

{\it Key words and phrases:}
Beta Negative Binomial Distribution,
Fisher Information Matrix,
Negative Binomial Distribution, ZINB, ZIBNB

\section{Introduction}
\label{sec:intro}

In the mathematical literature, $\Gamma(x) = \int_0^\infty t^{x-1} e^{-t} dt$, $x>0$ stands for the {\it gamma} function, $\Psi(\cdot) = \Gamma'(\cdot) / \Gamma(\cdot)$ is called the {\it digamma} function (see, for example, Section~6.3 in \cite{abramowitz1964handbook}), and $\Psi_1(\cdot) = \Psi'(\cdot)$ is called the {\it trigamma} function (see, for example, Section~6.4 in \cite{abramowitz1964handbook}). Except for some special values of $x$, these functions have to be calculated numerically and approximately.

In the statistical literature, the trigamma function $\Psi_1(\cdot)$ naturally shows up in calculating the Fisher information matrices of binomial or negative binomial related distributions. Quantities in the form of $E\Psi_1(\nu+Y)$ with $Y$ following a binomial, negative binomial, or other related distribution need to be calculated for the relevant Fisher information matrices. For examples, \cite{guo2020} considered computing $E\Psi_1(\nu + Y)$ with $Y$ following a negative binomial (NB) distribution; \citeauthor{aldirawi2022modeling} (2022, Example~3) needed to compute a similar quantity for zero-altered negative binomial (ZANB) or negative binomial hurdle (NBH) distributions; and  \cite{dousti2022r} computed expectations in the form of $E\Psi_1(r + \alpha + \beta + Y)$ with $Y$ following a beta negative binomial (BNB) distribution. 

The computation of $E{\Psi_1 (\nu + Y)}$ with $Y\sim{\rm NB}(\nu, p)$, $\nu>0$ and $p\in (0,1)$ needed for the Fisher information matrix has been investigated as a challenging problem in the literature. Due to the lack of an efficient algorithm, researchers have therefore turned to advocate an approximation known as the observed/empirical Fisher information \citep{adamidis1999theory, hilbe2011negative, chen2020statistical} to replace it in practice. It has been noted that the observed/empirical Fisher information may not provide a reliable approximation of the true Fisher information even with a large sample size \citep{efron1978assessing, kunstner2019limitations}. To address this issue, \cite{guo2020} proposed a numerical method to compute $E{\Psi_1 (\nu + Y)}$ approximately, which we call {\it GFWL's} method according to the initial letters of the authors' last names. More specifically, for $Y\sim{\rm NB}(\nu, p)$ and a large enough integer $M$, GFWL's method utilizes the following approximation
\begin{equation}\label{eq:Guo2020}
E\Psi_1 (\nu + Y) \approx \sum_{k=0}^M \Psi_1(k+\nu) P(Y=k) + \frac{1}{2} \Psi_1(M+1+\nu) P(Y > M)
\end{equation}
with an upper bound of error $\frac{1}{2} \Psi_1(M+1+\nu) P(Y > M)$. Note that GFWL's method needs to calculate the trigamma function $M+2$ times, whose computational time and accuracy level not only depend on $M$ but also the numerical approximation of the trigamma function.

For other types of  distributions of $Y$, such as ${\rm BNB}(r, \alpha, \beta)$,  \cite{dousti2022r} used a more traditional approach, call a {\it Monte Carlo} estimate 
\begin{equation}\label{eq:monte_carlo}
E\Psi_1 (\nu + Y) \approx \frac{1}{M} \sum_{i=1}^M \Psi_1(\nu + Y_i)
\end{equation}
where $Y_1, \ldots, Y_M$ are randomly sampled from ${\rm BNB}(r, \alpha, \beta)$. It is known that the accuracy level of a Monte Carlo estimator is of $o_P(M^{-1/2})$ (see, for example, \cite{dasgupta2008asymptotic}).

For regression analysis involving regular, zero-inflated, or zero-altered binomial, beta binomial, negative binomial, or beta negative binomial models, we need to calculate quantities such as 
\[
\sum_{i=1}^n c_i E\Psi_1 (\nu_i + Y_i)
\]
as an important component in the Fisher information matrix, where $n$ is the sample size, and $Y_i$, for example, follows ${\rm BNB}(r_i, \alpha_i, \beta_i)$.
On one hand, we need to calculate each $E\Psi_1 (\nu_i + Y_i)$ quickly since we need to calculate $n$ of them. On the other hand, since the inverse of the Fisher Information matrix needs to be calculated to obtain the corresponding confidence intervals of the model parameters, we also need to calculate it in a precise way so that the resulting inverse Fisher information matrix is stable and accurate.
These are the major motivations of ours to calculate $E\Psi_1(\nu+Y)$ as efficient and precise as needed.

In this paper, for fairly general non-negative integer-valued $Y$, we develop in Section~\ref{sec:calculate_E_Psi_1} a new approximation 
\begin{equation}\label{eq:trigamma_free}
E\Psi_1(\nu+Y) \approx \Psi_1(\nu) - \sum_{y=0}^M \frac{P(Y>y)}{(\nu+y)^2}    
\end{equation}
We call it a {\it trigamma-free} approach since it only needs to calculate the trigamma function once at $\nu > 0$ regardless of $M$. We show by numerical studies in Section~\ref{sec:verify} that our approximation~\eqref{eq:trigamma_free} is faster and more accurate with large $M$. For moderate or small $M$, we develop in Section~\ref{sec:improved_approximation} a modified approximation  
\begin{equation}\label{eq:improved_trigamma_free}
E\Psi_1(\nu+Y) \approx \Psi_1(\nu) - \sum_{y=0}^M \frac{P(Y>y)}{(\nu+y)^2} - \rho_* \cdot \frac{P(Y > M+1)}{\nu+M}       
\end{equation}
where $\rho_* = \frac{1}{2} + \frac{\nu+M}{2(\nu+M+1)(\nu+M+2)}$, which is as accurate as, but faster in many cases than, GFWL's method. We apply our trigamma-free approximation \eqref{eq:trigamma_free} with large enough $M$ for calculating Fisher information matrices for probabilistic models in Section~\ref{sec:calculate_Fisher_information} and regression models in Section~\ref{sec:application_ZI_regression_model}.

\section{Methods}
\label{sec:meth}

In this section, we propose two approximations for $E\Psi_1(\nu+Y)$ given an integer $M>0$.

\subsection{Calculating $E\Psi_1(\nu+Y)$ for non-negative integer-valued $Y$}\label{sec:calculate_E_Psi_1}

In this paper, we consider a fairly general discrete random variable $Y$, which takes non-negative integer values. Examples not only include binomial, negative binomial, beta binomial, beta negative binomial distributions, but also their zero-inflated and hurdle versions (see, for example, \cite{dousti2022r}).

According to, for example, Formula~6.3.5 in \cite{abramowitz1964handbook}, we have
\begin{eqnarray*}
\Psi(x + 1) - \Psi(x) &=& \frac{1}{x}\\
\Psi_1(x + 1) - \Psi_1(x) &=& -\frac{1}{x^2}
\end{eqnarray*}
for any $x>0$. Then for any positive integer $y$, 
\begin{eqnarray}
\Psi(x + y) - \Psi(x) &=& \sum_{j=0}^{y-1} \frac{1}{x+j}\label{eq:Psi_x+y}\\
\Psi_1(x + y) - \Psi_1(x) &=& -\sum_{j=0}^{y-1} \frac{1}{(x+j)^2}\label{eq:Psi_x+y_Psi_x}
\end{eqnarray}
Based on \eqref{eq:Psi_x+y} and \eqref{eq:Psi_x+y_Psi_x}, we obtain the following explicit expressions of expectations of digamma or trigamma functions. The detailed proof, as well as other proofs, has been relegated to the Appendix.

\begin{theorem}\label{thm:E_Psi_1}
If $Y$ takes only non-negative integer values, then for any $\nu>0$,
\begin{eqnarray}
E\Psi(\nu +Y) &=& \Psi(\nu) + \sum_{y=0}^\infty \frac{P(Y > y)}{\nu + y}\nonumber\\
E\Psi_1(\nu +Y) &=& \Psi_1(\nu) - \sum_{y=0}^\infty \frac{P(Y > y)}{(\nu + y)^2}\label{eq:E_Psi_1}
\end{eqnarray}
\end{theorem}

For bounded $Y$, such as binomial, beta binomial, and their zero-inflated or zero-altered distributions, their expectations can be calculated directly due to the following corollary. Actually, if $Y$ takes value in $\{0, 1, \ldots, n\}$ only, then $P(Y>y)=0$ for all $y\geq n$, and we have Corollary~\ref{col:E_Psi_1_n} below as a direct conclusion of Theorem~\ref{thm:E_Psi_1}. 
For these kinds of distributions, we do not need approximations for $E\Psi(\nu+Y)$ or $E\Psi_1(\nu+Y)$, since we can calculate them directly.

\begin{corollary}\label{col:E_Psi_1_n}
If $Y$ takes value in $\{0, 1, \ldots, n\}$ only, then for any $\nu>0$,
\begin{eqnarray*}
E\Psi(\nu+Y) &=& \Psi(\nu) + \sum_{y=0}^{n-1}\frac{P(Y>y)}{\nu+y}\\
E\Psi_1(\nu+Y) &=& \Psi_1(\nu) - \sum_{y=0}^{n-1}\frac{P(Y>y)}{(\nu+y)^2}
\end{eqnarray*}
\end{corollary}

Our main goal in this paper is to find an efficient and accurate approximation for $E\Psi_1(\nu +Y)$ when $Y$ is unbounded.  Theorem~\ref{thm:E_Psi_1} provides a natural approximation for $E\Psi_1(\nu +Y)$, denoted by 
\begin{equation}\label{eq:Psi_1_aproximation_general}
\Psi_e(\nu, M, 0) = \Psi_1(\nu) - \sum_{y=0}^M \frac{P(Y > y)}{(\nu + y)^2}
\end{equation}
with a pre-determined positive integer $M$.  According to the following theorem, the error of the approximation $\Psi_e(\nu, M, 0)$ is bounded by $P(Y>M+1)/(\nu+M)$, which can be arbitrarily small by choosing a large enough $M$.

\begin{theorem}\label{thm:Psi_approximation_general}
If $Y$ takes non-negative integer values only, then for any real number $\nu>0$ and any integer $M\geq 0$,
\begin{equation}\label{eq:approx_1_bound}
\left|E\Psi_1(\nu +Y) - \Psi_1(\nu) + \sum_{y=0}^M \frac{P(Y > y)}{(\nu + y)^2}\right| \leq \frac{P(Y > M+1)}{\nu + M}
\end{equation}
\end{theorem}

\subsection{An alternative approximation}\label{sec:improved_approximation}

In Section~\ref{sec:calculate_E_Psi_1}, we propose $\Psi_e(\nu, M, 0)$ as described in \eqref{eq:Psi_1_aproximation_general} for approximating $E\Psi_1(\nu +Y)$ with an upper bound $P(Y>M+1)/(\nu+M)$ for the error term $\sum_{y=M+1}^\infty \frac{P(Y > y)}{(\nu + y)^2}$. In this section, we develop an alternative approximation which reduces the upper bound by a half.

According to \eqref{eq:E_Psi_1} and \eqref{eq:approx_1_bound},  
\[
0\leq \Psi_e(\nu, M, 0) - E\Psi_1(\nu +Y) =  \sum_{y=M+1}^\infty \frac{P(Y > y)}{(\nu + y)^2} \leq \frac{P(Y > M+1)}{\nu + M}
\]
Then a calibrated approximation for $E\Psi_1(\nu +Y)$ could be defined as
\begin{equation}\label{eq:Psi_1_aproximation_general_epsilon}
\Psi_e(\nu, M, \rho) = \Psi_1(\nu) - \sum_{y=0}^M \frac{P(Y > y)}{(\nu + y)^2} - \rho \cdot \frac{P(Y > M+1)}{\nu + M}
\end{equation}
where $\rho\in [0,1]$.
To find the best $\rho$ for $\Psi_e(\nu, M, \rho)$, we need the following lemma which provides a better range for the error term $\sum_{y=M+1}^\infty \frac{P(Y > y)}{(\nu + y)^2}$ of $\Psi_e(\nu, M, 0)$.

\begin{lemma}\label{lem:lower_bound_difference}
\begin{eqnarray*}
\sum_{y=M+1}^\infty \frac{P(Y > y)}{(\nu + y)^2} &\geq & \frac{P(Y>M+1)}{\nu+M+1} - \sum_{k=M+2}^\infty \frac{P(Y=k)}{\nu+k}\\
\sum_{y=M+1}^\infty \frac{P(Y > y)}{(\nu + y)^2} &\leq & \frac{P(Y>M+1)}{\nu+M} - \sum_{k=M+2}^\infty \frac{P(Y=k)}{\nu + k-1}
\end{eqnarray*}
where ``$=$'' is true if and only if $P(Y>M+1)=0$.
\end{lemma}

With the aid of Lemma~\ref{lem:lower_bound_difference}, we obtain an upper bound for the error term of $\Psi_e(\nu, M, \rho)$ and derive the best $\rho$ corresponding to the smallest upper bound as described in the following theorem.

\begin{theorem}\label{thm:rho=1/2} 
An upper bound for $|\Psi_e(\nu, M, \rho) - E\Psi_1(\nu+Y)|$ is
\[
\frac{P(Y>M+1)}{\nu+M} \cdot \max\left\{|1-\rho|, \left|\rho - \frac{\nu+M}{(\nu+M+1)(\nu+M+2)}\right|\right\}
\]
which is minimized at 
$\rho_*= \frac{1}{2} + \frac{\nu+M}{2(\nu+M+1)(\nu+M+2)}$. 
\hfill{$\Box$}
\end{theorem}

According to Theorem~\ref{thm:rho=1/2}, $\Psi_e(\nu, M, \rho_*)$ is an alternative approximation for $E\Psi_1(\nu+Y)$.

\section{Numerical Studies}
\label{sec:verify}

In this section, we use numerical studies to compare the proposed approximations $\Psi_e(\nu, M, 0)$ and $\Psi_e(\nu, M, \rho_*)$ with the Monte Carlo estimator \eqref{eq:monte_carlo} and the GFWL's method \eqref{eq:Guo2020} for negative binomial distribution, which has been used as a common non-negative integer-valued probabilistic model \citep{cardinal1999application, latour1998existence, zhu2011negative, zhu2012zero}. 

Given a negative binomial distribution with parameters $\boldsymbol\theta = (\nu, p)^T \in (0, \infty)$ $\times (0,1)$ (see, for example, \cite{wang2020identifying}), its probability mass function (pmf) is given by
\[
f_{\boldsymbol\theta}(y) = \frac{\Gamma(\nu+y)}{\Gamma(y+1) \Gamma(\nu)} p^{\nu} (1-p)^y,
\]
with $y \in \{0,1,\ldots\}$,  mean value $\nu(1-p)/p$ and variance $\nu(1-p)/p^2$. 

\subsection{Accuracy of approximations}\label{sec:accuracy_approximation}

To compare the accuracy of different approximations, we adopt $\Psi_e(\nu, 10^6, 0)$ as the exact value of $E\Psi_1(\nu+Y)$, whose error term is below $P(Y>10^6+1)/(\nu+10^6)$ according to Theorem~\ref{thm:Psi_approximation_general}. For all $(\nu, p)$'s used in this section, the error term is less than $-9511$ in log-scale, which is essentially a numerical zero.

\begin{figure}[ht]
    \centering
    \includegraphics[scale=0.45]{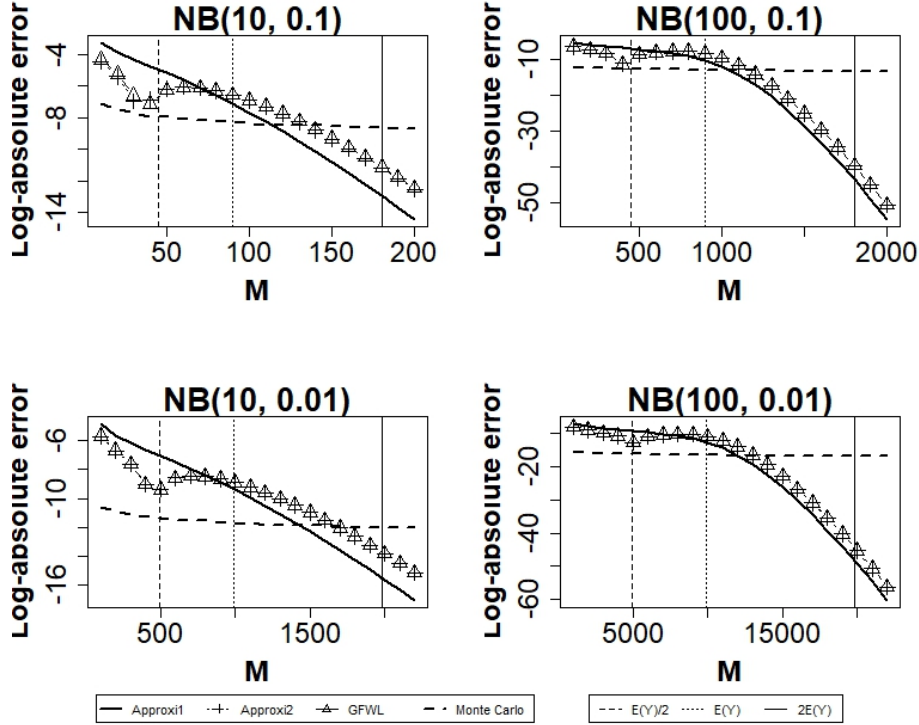}
    \caption{Log-scale absolute errors of approximations $\Psi_e(\nu, M, 0)$ (Approxi1), $\Psi_e(\nu, M, \rho_*)$ (Approxi2), GFWL's method, and (median) Monte Carlo estimates for various $M$ and  ${\rm NB}(\nu, p)$ compared with $\Psi_e(\nu, 10^6, 0)$, as well as vertical lines $E(Y)/2=\nu(1-p)/(2p)$, $E(Y)$ and $2E(Y)$
    }
\label{fig:NB_4MethodsComparison_2by2_log}
\end{figure}

We display in log-scale the absolute errors of different approximations for $E\Psi_1(\nu+Y)$ along with an increasing $M$ in Figure~\ref{fig:NB_4MethodsComparison_2by2_log} for $Y\sim {\rm NB}(\nu, p)$ with various $(\nu, p)$.
Since the Monte Carlo estimates are random, for each ${\rm NB}(\nu, p)$ and each $M$, we obtain 1,000 independent Monte Carlo estimates and calculate the median absolute error of the Monte Carlo estimates in log-scale, which are plotted in dash lines in Figure~\ref{fig:NB_4MethodsComparison_2by2_log}.

From Figure~\ref{fig:NB_4MethodsComparison_2by2_log}, we can see that {\it (i)} $\Psi_e(\nu, M, \rho_*)$ and the GFWL's method provide nearly the same approximations; {\it (ii)} the approximations based on $\Psi_e(\nu, M, 0)$, $\Psi_e(\nu, M, \rho_*)$ and the GFWL's method converge to $\Psi_e(\nu, 10^6, 0)$'s (regarded as the exact value) much faster than the Monte Carlo estimator as $M$ increases; {\it (iii)} it seems that $\Psi_e(\nu, M, \rho_*)$ and the GFWL's method are better than $\Psi_e(\nu, M, 0)$ if $M < E(Y)/2$ (dash vertical lines in Figure~\ref{fig:NB_4MethodsComparison_2by2_log}); {\it (iv)} it seems that $\Psi_e(\nu, M, 0)$ is more accurate than $\Psi_e(\nu, M, \rho_*)$ and the GFWL's method if $M \geq E(Y)$ (dotted vertical lines in Figure~\ref{fig:NB_4MethodsComparison_2by2_log}); {\it (v)} it seems that an $M > 2E(Y)$ (solid vertical lines in Figure~\ref{fig:NB_4MethodsComparison_2by2_log}) is large enough for $\Psi_e(\nu, M, 0)$, $\Psi_e(\nu, M, \rho_*)$ and the GFWL's method to obtain much better approximations than the Monte Carlo estimator.  

It should be noted that compared with the GFWL's method, the improvement of the approximation $\Psi_e(\nu, M, 0)$ is significant in terms of approximation errors. For example, in Table~\ref{tab:rho_guo_NB_absolute_error}, we list the log-scale absolute errors of $\Psi_e(\nu, M, 0)$ ($\log|\epsilon_1| = \log|\Psi_e(\nu, M, 0) - E\Psi_1(\nu+Y)|$, $Y\sim $ NB(100,0.01)) and GFWL's ($\log|\epsilon_g|$). One can see that as $M$ increases from 10,000 (about $E(Y)$) to 22,000 (about $2E(Y)$), the ratio ($|\epsilon_g/\epsilon_1|$) of absolute errors increases rapidly from $6.24$ to $62.04$.

\begin{table}[ht]
\centering
\footnotesize
\begin{tabular}{|l|rrrrrrr|}\hline
$M$  &  10000  &  12000  &  14000  &  16000  &  18000  &  20000  &  22000\\ \hline
$\Psi_e(\nu, M, 0)$ ($\log|\epsilon_1|$) & -12.70 & -16.61 & -22.54 & -30.17 & -39.17 & -49.28 & -60.32\\ \hline
GFWL's ($\log|\epsilon_g|$)  & -10.87  & -13.97  & -19.40  & -26.68  & -35.42  & -45.33  & -56.19\\ \hline
Log Ratio ($\log|\epsilon_g/\epsilon_1|$) &      1.83 &   2.64 &   3.14 &   3.49 &   3.75 &   3.95 &   4.13\\ \hline
Ratio ($|\epsilon_g/\epsilon_1|$)  &   6.24  &  14.02  &  23.13  &  32.72  &  42.32  &  52.15  &  62.04\\ \hline
\end{tabular}
\normalsize
\caption{Log-scale Absolute Errors and Ratios of Approximations $\Psi_e(\nu, M, 0)$ and GFWL's for $E\Psi_1(\nu + Y)$ with $Y\sim {\rm NB}(100,0.01)$ 
}
\label{tab:rho_guo_NB_absolute_error}
\end{table}

\subsection{Sensitivity to model parameters}\label{sec:sensitivity_approximation}

In this section, we conduct a simulation study to compare the sensitivities of different approximations to model parameters. 

We adopt NB$(10, 0.1)$ as the baseline model for illustration purposes. According to Section~\ref{sec:accuracy_approximation}, $\Psi_e(\nu,10^6,0)$ provides a numerically precise approximation $0.01104294$ for $E\Psi_1(\nu+Y)$ with $Y\sim {\rm NB}(10,0.1)$. If the model parameters $\nu$ and $p$ are not given precisely but, say, estimated from some data, we want to know if the approximations based on NB$(\hat\nu, \hat{p})$ are still good for approximating $E\Psi_1(\nu+Y)$.

For illustration purposes, we simulate $B=1000$ random samples from NB$(10,0.1)$, each with the sample size $N=1000$. From the $i$th random sample, we use the R function {\tt new.mle} in package {\tt AZIAD} \citep{dousti2022r} to find the maximum likelihood estimates $(\hat\nu_i, \hat{p}_i)$ for the negative binomial model. For readers' reference, $(9.119, 11.078)$ and $(0.092, 0.110)$ are the pairs of $(0.025, 0.975)$th quantiles of $\hat\nu_i$'s and $\hat{p}_i$'s, respectively, which roughly describe the ranges of $\hat\nu_i$'s and $\hat{p}_i$'s.

\begin{figure}[ht]
    \centering
    \includegraphics[scale=0.5]{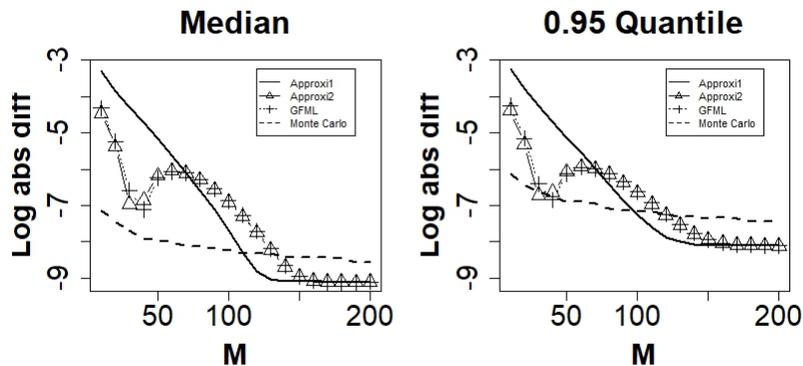}
    \caption{Medians and $0.95$th quantiles of log-scale absolute differences between the approximations based on the mle $(\hat\nu_i, \hat{p}_i)$'s and the precise value $\Psi_e(\nu,10^6,0)$ for $E\Psi_1(\nu+Y)$ with $Y\sim {\rm NB}(10,0.1)$
    }
\label{fig:NB_3MethodsComparison_log_absolute_error}
\end{figure}

In Figure~\ref{fig:NB_3MethodsComparison_log_absolute_error}, for each $M$ in $\{10, 20, \ldots, 200\}$, we display the medians and $0.95$th quantiles of the 1000 log-scale absolute differences between the approximations based on the estimates $(\hat\nu_i, \hat{p}_i)$ and the precise value $E\Psi_1(\nu+Y)$ based on $\Psi_e(\nu,10^6,0)$. The patterns are fairly similarly to Figure~\ref{fig:NB_4MethodsComparison_2by2_log} except that the three approximations,  $\Psi_e(\nu, M, 0)$ (Approxi1), $\Psi_e(\nu, M, \rho_*)$ (Approxi2) and GFWL's, are nearly the same when $M > 2E(Y)$, which is $180$ for NB$(10,0.1)$. It is because all the three approximations are so accurate that the corresponding median and $0.95$th quantile of the log absolute differences shown in Figure~\ref{fig:NB_3MethodsComparison_log_absolute_error} are mainly due to the variation of $\hat\nu_i$ and $\hat{p}_i$'s. On the contrary, when $M>2E(Y)$, the approximations based on Monte Carlo estimators \eqref{eq:monte_carlo} contain more variations than the other three approximations.

\subsection{Computational time}\label{sec:computational_time}

In terms of computation time, we provide the average time costs over $100$ repetitions for the three approximations across various ${\rm NB}(\nu, p)$ and $M$ in Table~\ref{tab:rho_guo_NB_computation_time}. The operations are timed in R on a laptop computer with Windows 10 Pro, Intel(R) Core(TM) i7-9850H 2.60GHz CPU and 32GB RAM. Overall, $\Psi_e(\nu, M, 0)$ and $\Psi_e(\nu, M, \rho_*)$ cost comparable time. When $\nu$ is small, such as $\nu=10$, $\Psi_e(\nu, M, 0)$ and $\Psi_e(\nu, M, \rho_*)$ are slower than the GFWL's method. Further analysis (not shown here) indicates that it is mainly because the calculation of $P(Y>y), y=0, \ldots, M$ with small $\nu$ takes longer time.
When $\nu$ is moderate or large, say $\nu\geq 100$ (further analysis, not shown here, indicates that $\nu$ can be as small as $40$), $\Psi_e(\nu, M, 0)$ and $\Psi_e(\nu, M, \rho_*)$ are faster than the GFWL's. It should be noted that the value of parameter $p$ seems not to affect the computation time (not shown here).

\begin{table}[ht]
\centering
\footnotesize
\begin{tabular}{|cccccccc|}
\hline
\multicolumn{8}{|c|}{\textbf{NB(10, 0.1) Computation Time ($\times 10^{-2}$ seconds)}} \\ \hline
\multicolumn{1}{|c|}{M}   & \multicolumn{1}{c|}{10}      & \multicolumn{1}{c|}{$10^2$}     & \multicolumn{1}{c|}{$10^3$}      & \multicolumn{1}{c|}{$10^4$}      & \multicolumn{1}{c|}{$10^5$}      & \multicolumn{1}{c|}{$10^6$}      & $10^7$      \\ \hline
\multicolumn{1}{|c|}{\bf{$\Psi_e(\nu, M, 0)$}}   & \multicolumn{1}{c|}{0.01} & \multicolumn{1}{c|}{0.004} & \multicolumn{1}{c|}{0.03} & \multicolumn{1}{c|}{0.31} & \multicolumn{1}{c|}{2.81} & \multicolumn{1}{c|}{26.2} & 263 \\ \hline
\multicolumn{1}{|c|}{\bf{$\Psi_e(\nu, M, \rho_*)$}}   & \multicolumn{1}{c|}{0.02} & \multicolumn{1}{c|}{0.010} & \multicolumn{1}{c|}{0.04} & \multicolumn{1}{c|}{0.33} & \multicolumn{1}{c|}{2.87} & \multicolumn{1}{c|}{26.5} & 263 \\ \hline
\multicolumn{1}{|c|}{GFWL} & \multicolumn{1}{c|}{0.01} & \multicolumn{1}{c|}{0.004} & \multicolumn{1}{c|}{0.02} & \multicolumn{1}{c|}{0.18} & \multicolumn{1}{c|}{1.72} & \multicolumn{1}{c|}{15.9} & 162 \\ \hline
\hline
\multicolumn{8}{|c|}{\textbf{NB(100, 0.1) Computation Time ($\times 10^{-2}$ seconds)}} \\ \hline
\multicolumn{1}{|c|}{M}   & \multicolumn{1}{c|}{10}      & \multicolumn{1}{c|}{$10^2$}     & \multicolumn{1}{c|}{$10^3$}      & \multicolumn{1}{c|}{$10^4$}      & \multicolumn{1}{c|}{$10^5$}      & \multicolumn{1}{c|}{$10^6$}      & $10^7$      \\ \hline
\multicolumn{1}{|c|}{\bf{$\Psi_e(\nu, M, 0)$}}   & \multicolumn{1}{c|}{0.001} & \multicolumn{1}{c|}{0.003} & \multicolumn{1}{c|}{0.03} & \multicolumn{1}{c|}{0.21} & \multicolumn{1}{c|}{1.43} & \multicolumn{1}{c|}{12.1} & 125 \\ \hline
\multicolumn{1}{|c|}{\bf{$\Psi_e(\nu, M, \rho_*)$}}   & \multicolumn{1}{c|}{0.001} & \multicolumn{1}{c|}{0.003} & \multicolumn{1}{c|}{0.03} & \multicolumn{1}{c|}{0.21} & \multicolumn{1}{c|}{1.23} & \multicolumn{1}{c|}{11.6} & 124 \\ \hline
\multicolumn{1}{|c|}{GFWL} & \multicolumn{1}{c|}{0.001} & \multicolumn{1}{c|}{0.003} & \multicolumn{1}{c|}{0.02} & \multicolumn{1}{c|}{0.23} & \multicolumn{1}{c|}{1.77} & \multicolumn{1}{c|}{15.8} & 159 \\ \hline
\hline
\multicolumn{8}{|c|}{\textbf{NB(500, 0.1) Computation Time ($\times 10^{-2}$ seconds)}}   \\ \hline
\multicolumn{1}{|c|}{M}   & \multicolumn{1}{c|}{10}      & \multicolumn{1}{c|}{$10^2$}     & \multicolumn{1}{c|}{$10^3$}      & \multicolumn{1}{c|}{$10^4$}      & \multicolumn{1}{c|}{$10^5$}      & \multicolumn{1}{c|}{$10^6$}      & $10^7$      \\ \hline
\multicolumn{1}{|c|}{\bf{$\Psi_e(\nu, M, 0)$}}   & \multicolumn{1}{c|}{0.001} & \multicolumn{1}{c|}{0.002} & \multicolumn{1}{c|}{0.02} & \multicolumn{1}{c|}{0.22} & \multicolumn{1}{c|}{1.52} & \multicolumn{1}{c|}{12.3} & 126 \\ \hline
\multicolumn{1}{|c|}{\bf{$\Psi_e(\nu, M, \rho_*)$}}   & \multicolumn{1}{c|}{0.001} & \multicolumn{1}{c|}{0.003} & \multicolumn{1}{c|}{0.02} & \multicolumn{1}{c|}{0.25} & \multicolumn{1}{c|}{1.53} & \multicolumn{1}{c|}{12.3} & 126 \\ \hline
\multicolumn{1}{|c|}{GFWL} & \multicolumn{1}{c|}{0.001} & \multicolumn{1}{c|}{0.003} & \multicolumn{1}{c|}{0.02} & \multicolumn{1}{c|}{0.22} & \multicolumn{1}{c|}{1.75} & \multicolumn{1}{c|}{15.6} & 163 \\ \hline
\hline
\multicolumn{8}{|c|}{\textbf{NB(1000, 0.1) Computation Time ($\times 10^{-2}$ seconds)}}  \\ \hline
\multicolumn{1}{|c|}{M}   & \multicolumn{1}{c|}{10}      & \multicolumn{1}{c|}{$10^2$}     & \multicolumn{1}{c|}{$10^3$}      & \multicolumn{1}{c|}{$10^4$}      & \multicolumn{1}{c|}{$10^5$}      & \multicolumn{1}{c|}{$10^6$}      & $10^7$      \\ \hline
\multicolumn{1}{|c|}{\bf{$\Psi_e(\nu, M, 0)$}}   & \multicolumn{1}{c|}{0.001} & \multicolumn{1}{c|}{0.002} & \multicolumn{1}{c|}{0.02} & \multicolumn{1}{c|}{0.24} & \multicolumn{1}{c|}{1.51} & \multicolumn{1}{c|}{12.4} & 124 \\ \hline
\multicolumn{1}{|c|}{\bf{$\Psi_e(\nu, M, \rho_*)$}}   & \multicolumn{1}{c|}{0.001} & \multicolumn{1}{c|}{0.003} & \multicolumn{1}{c|}{0.02} & \multicolumn{1}{c|}{0.26} & \multicolumn{1}{c|}{1.53} & \multicolumn{1}{c|}{12.5} & 126 \\ \hline
\multicolumn{1}{|c|}{GFWL} & \multicolumn{1}{c|}{0.001} & \multicolumn{1}{c|}{0.003} & \multicolumn{1}{c|}{0.02} & \multicolumn{1}{c|}{0.22} & \multicolumn{1}{c|}{1.77} & \multicolumn{1}{c|}{15.8} & 162 \\ \hline
\end{tabular}
\normalsize
\caption{Average Computational Time for $\Psi_e(\nu, M, 0)$, $\Psi_e(\nu, M, \rho_*)$ and GFWL's 
}
\label{tab:rho_guo_NB_computation_time}
\end{table}

As a conclusion, we recommend $\Psi_e(\nu, M, 0)$ as described in \eqref{eq:Psi_1_aproximation_general} with $M > 2E(Y)$ for approximating $E\Psi_1(\nu+Y)$, whose error term is bounded by $P(Y>M+1)/(\nu+M)$. Note that if $M$ is not large enough, such as $M<E(Y)/2$, then $\Psi_e(\nu, M, \rho_*)$ could be more accurate than $\Psi_e(\nu, M, 0)$.

\section{Applications}
\label{sec:applications}

In this section, we provide two kinds of applications of $\Psi_e(\nu, M, 0)$ for approximating $E\Psi_1(\nu + Y)$, called the trigamma-free approach or method. One is for obtaining the Fisher information matrix and confidence intervals for probabilistic model parameters. The other one is for regression model parameters. 

For both applications, the Monte Carlo estimator~\eqref{eq:monte_carlo} has been commonly used in practice. We will show how much one may gain if our trigamma-free method is used instead. We skip $\Psi_e(\nu, M, \rho_*)$ and GFWL's methods in this section since $\Psi_e(\nu, M, 0)$ is the one that we recommend after comprehensive simulation studies in Section~\ref{sec:verify}.

\subsection{Applications to probabilistic models}\label{sec:calculate_Fisher_information}

In this section, we use both a simulated example and a real dataset to show the applications of the recommended approximation $\Psi_e(\nu, M, 0)$ \eqref{eq:Psi_1_aproximation_general} for probabilistic models.

\begin{example}\label{ex:ZANB_ZINB}\label{ex:ZANB}
{\bf Zero-altered negative binomial (ZANB) or zero-inflated negative binomial (ZINB) model}\quad 
{\rm
To calculate the Fisher information matrix of a ZANB distribution (see, for example, Example~3 in \cite{aldirawi2022modeling}) or a ZINB distribution (see, for example, Example~2.5 in \cite{dousti2022r}, where $r$ was used instead of $\nu$), $E\Psi_1(\nu+Y)$ needs to be calculated with $Y\sim{\rm NB}(\nu, p)$ given $\nu>0$ and $p\in (0,1)$.
A Monte Carlo estimator \eqref{eq:monte_carlo} was used by \cite{dousti2022r} to estimate $E\Psi_1(\nu+Y)$ and then the Fisher information matrix of the corresponding ZINB distribution. 

We conduct a simulation study here to compare the Monte Carlo estimator with our trigamma-free estimator $\Psi_e(\nu, M, 0)$ \eqref{eq:Psi_1_aproximation_general}. 
For illustration purposes, we consider the ZINB model with parameters $\phi=0.4$, $\nu=10$, and $p=0.1$.  We simulate $B=1000$ independent datasets with sample size $N=1000$ from ${\rm ZINB}(\phi, \nu, p)$ and denote the datasets as ${\mathbf D}_i$, $i=1, \ldots, B$. 

For each dataset ${\mathbf D}_i$, we first use the R package {\tt AZIAD} \citep{dousti2022r} to obtain the maximum likelihood estimate (MLE) $(\hat{\phi}_i, \hat\nu_i, \hat{p}_i)$  of the ZINB distribution. To calculate the Fisher information matrix at $(\hat{\phi}_i, \hat\nu_i, \hat{p}_i)$, we follow the formula in Example~2.5 of \cite{dousti2022r}, which contains $E\Psi_1(\hat{\nu}_i+Y)$ with $Y\sim {\rm NB}(\hat{\nu}_i, \hat{p}_i)$. We use either the Monte Carlo estimator \eqref{eq:monte_carlo} with sample size $M$ or our trigamma-free approximation $\Psi_e(\hat\nu_i, M, 0)$ to estimate $E\Psi_1(\hat{\nu}_i+Y)$ and denote the resulting Fisher information matrix as $\hat{\mathbf F}_{iM}$. 
To evaluate the accuracy of $\hat{\mathbf F}_{iM}$, we use $\Psi_e(\hat\nu_i, 10^6, 0)$ to replace $E\Psi_1(\hat\nu_i + Y)$, denote the corresponding Fisher information matrix as ${\mathbf G}_{i}$, and treat ${\mathbf G}_i$ as the exact Fisher information matrix.
Note that one may also denote ${\mathbf G}$ as the Fisher information matrix at $(\phi, \nu, p)$, which is not the target in this example since $(\phi, \nu, p)$ are typically unknown in practice.
We then use two kinds of measures to check the differences between $\hat{\mathbf F}_{iM}^{-1}$ and ${\mathbf G}_i^{-1}$, the Frobenius norm \citep{datta2010numerical} (or the Hilbert-Schmidt norm \citep{akhiezer2013theory})  and the maximum length change of the resulting confidence intervals \citep{gentle2012numerical}.

More specifically, given a real matrix ${\mathbf A}=(a_{ij})_{ij}$, its Frobenius norm $\lVert A \rVert_{F} = \sqrt{\sum_i \sum_j a_{ij}^2}$. Based on that, we define the Frobenius distance between the two inverse Fisher information matrices, 
$\lVert \hat{\mathbf F}_{iM}^{-1} - {\mathbf G}_i^{-1} \Vert_{F}$.
On the other hand, the maximum length change of confidence intervals for these two inverse Fisher information matrices is defined as $\max_{k}|\sqrt{b_{kk}} - \sqrt{d_{kk}}|$, where $b_{kk}$ and $d_{kk}$ are the $k$th diagonal elements of $\hat{\mathbf F}_{iM}^{-1}$ and ${\mathbf G}_i^{-1}$, respectively.

In Table~\ref{tab:ZINB_Fnorm_MaxLengthChangeCI_MK}, we provide the average distances between the estimated and the exact Fisher information matrices over $B=1000$ simulated datasets across different $M$'s. 
One can see that 
our trigamma-free method obtains essentially the exact Fisher information matrix when $M\geq 1000$, while the Monte Carlo estimate approaches the exact Fisher information matrix slowly as the sample size $M$ increases. 

\begin{table}[ht]
\centering
\begin{tabular}{|c|c|c|c|}
\hline
{\bf Method} & $M$ & {\bf Frobenius Norm} & {\bf Max Length Change} \\ \hline
 & 100    & 2.01                   & 0.12                        \\ 
Trigamma-free  & 1000   & 0                       & 0                                \\ 
  & 5000  & 0                       & 0                                \\ 
  & 20000 & 0                       & 0                                \\ \hline
  & 1000   & 9.54                   & 0.24                        \\ 
Monte Carlo   & 5000   & 4.21                   & 0.10                        \\  
  & 20000  & 2.06                   & 0.05                        \\ \hline
\end{tabular}
\caption{Average Distance over 1000 Simulations between the Estimated and Exact Inverse Fisher Information Matrices for ZINB Distribution with $\phi=0.4$, $\nu=10$ and $p=0.1$}
\label{tab:ZINB_Fnorm_MaxLengthChangeCI_MK}
\end{table}

For practical purposes, we also compare the empirical coverage rates of the $95\%$ confidence intervals (see also \cite{dousti2022r}) based on the estimated Fisher information matrix, that is, the proportion out of 1000 simulations that the obtained confidence intervals cover the true parameter values.  
In Table~\ref{tab:ZINB_CI_CoverRate_MK}, we provide the coverage rates for our trigamma-free method and the Monte Carlo estimator across different $M$.
One can see that the coverage rates obtained by our method for $\phi$, $\nu$ and $p$ are stable at $0.950$, $0.954$ and $0.950$, respectively, when $M$ is as small as $150$, while the Monte Carlo estimator achieves about the same numbers with $M\geq 5000$. In other words, in terms of confidence intervals, our trigamma-free approximation with $M=150$ provides consistent results with the ones based on the ``exact'' Fisher information matrix, while the Monte Carlo estimator needs $M = 5000$ or higher to achieve comparable results. Note that the empirical coverage rates for $\nu$ are different from the nominal rate $95\%$, which is a theoretical coverage rate when the sample size $N$ goes to infinity.

\begin{table}[ht]
\centering
\begin{tabular}{|c|cccc|ccc|}
\hline
                       
{\bf Method} & \multicolumn{4}{c|}{\textbf{Trigamma-free}}                                              & \multicolumn{3}{c|}{\textbf{Monte Carlo}}                                              \\ \hline
$M$ & \multicolumn{1}{c|}{\textbf{100}} & \multicolumn{1}{c|}{\textbf{150}} & \multicolumn{1}{c|}{\textbf{1000}} & \textbf{5000} & \multicolumn{1}{c|}{\textbf{1000}} & \multicolumn{1}{c|}{\textbf{5000}} & \textbf{20000} \\ \hline
$\phi$ & \multicolumn{1}{c|}{0.950}          & \multicolumn{1}{c|}{0.950}           & \multicolumn{1}{c|}{0.950}           & 0.950            & \multicolumn{1}{c|}{0.950}          & \multicolumn{1}{c|}{0.950}          & 0.950           \\ %
$\nu$  & \multicolumn{1}{c|}{0.967}         & \multicolumn{1}{c|}{0.954}         & \multicolumn{1}{c|}{0.954}          & 0.954           & \multicolumn{1}{c|}{0.957}         & \multicolumn{1}{c|}{0.954}         & 0.955          \\ 
$p$   & \multicolumn{1}{c|}{0.969}          & \multicolumn{1}{c|}{0.950}          & \multicolumn{1}{c|}{0.950}           & 0.950            & \multicolumn{1}{c|}{0.954}         & \multicolumn{1}{c|}{0.949}         & 0.950           \\ \hline
\end{tabular}
\caption{Coverage Rates of $95\%$ Confidence Intervals over 1000 Simulations for ZINB Distribution with $\phi=0.4$, $\nu=10$ and $p=0.1$}
\label{tab:ZINB_CI_CoverRate_MK}
\end{table}

In terms of the computational time, 
it takes $1.0\times 10^{-4}$ seconds for our trigamma-free method with $M=150$. As for the Monte Carlo estimator, it needs 
$1.8\times10^{-3}$ and $9.6\times10^{-3}$ seconds for $M=5000$ and $20,000$, respectively. 
}\hfill{$\Box$}
\end{example}

\begin{example}\label{ex:ZABNB_ZIBNB}{\bf Zero-altered beta negative binomial (ZABNB) or zero-inflated beta negative binomial (ZIBNB)}\quad {\rm
To calculate the Fisher information matrix of a ZABNB distribution (see, for example, Example~2.3 in \cite{dousti2022r}) or a ZIBNB distribution (see, for example, Example~S.2 in the Supplementary Materials of \cite{dousti2022r}, where $r$ was used instead of $\nu$), $E\Psi_1(\nu+Y)$, $E\Psi_1(\nu+\alpha+\beta+Y)$, and $E\Psi_1(\beta+Y)$ need to be calculated with $\nu>0$, $\alpha>0$, $\beta>0$, and $Y\sim{\rm BNB}(\nu, \alpha, \beta)$.

In this example, we consider a real dataset, {\tt DebTrivedi} from the R package {\tt MixAll}~\citep{mixall}. Following \cite{dousti2022r}, we focus on the variable {\tt ofp}, the number of physician office visits. According to Section~3.5 in \cite{dousti2022r}, the ZIBNB model and beta negative binomial hurdle (BNBH or ZABNB) model are significantly better than other probabilistic models. In this example, we focus on ZIBNB for illustration purposes.

By setting a random seed $323$ in R and employing the R package {\tt AZIAD} \citep{dousti2022r} with initial values of $\nu=3$, $\alpha=4$ and $\beta=3$, we obtain the MLE $(\hat{\phi}=0.094, \hat\nu=4.733, \hat\alpha=4.504, \hat\beta=4.733)$ for {\tt ofp} under ZIBNB model. Similar as in Example~\ref{ex:ZANB_ZINB}, we use our trigamma-free method and the Monte Carlo estimator to calculate the Fisher information matrix $\hat{\mathbf F}_{M}$ with various $M$ at $(\hat\phi, \hat\nu, \hat\alpha, \hat\beta)$, as well as ${\mathbf G}$ based on our trigamma-free method with $M=10^6$ and treated as the exact Fisher information matrix. We also calculate the Frobenius distance and the maximum length change of the resulting confidence intervals between $\hat{\mathbf F}_M^{-1}$ and ${\mathbf G}^{-1}$, which are provided in Table~\ref{tab:ZIBNB_trigammafree_DebTrivedi_ofp_twodistances}.

\begin{table}[ht]
\centering
\begin{tabular}{|c|c|c|c|}
\hline
{\bf Method} & $M$ & \textbf{Frobenius Norm} & \textbf{Max Length Change} \\ \hline
 & 5000  & $1.32 \times 10^{14}$               & $3.63 \times 10^6$                       \\ 
Trigamma-free & 10000 & $4.71 \times 10^{12}$               & $1.64 \times 10^5$                         \\ 
 & 11000 & 0                       & 0                                \\
 & 12000 & 0                       & 0                                \\ \hline
 & 
5000  & $1.05 \times 10^{14}$               & $7.26 \times 10^6$                          \\ 
Monte Carlo  & 10000 & $1.05 \times 10^{14}$               & $7.26 \times 10^6$                          \\ 
  & $10^{5}$  & $1.05 \times 10^{14}$               & $7.26 \times 10^6$                            \\ 
  & $10^{6}$  & $1.05 \times 10^{14}$               & $7.26 \times 10^6$                            \\ \hline 
\end{tabular}
\caption{Distance between the Estimated and Exact Inverse Fisher Information Matrices for 
{\tt ofp} in {\tt DebTrivedi} under ZIBNB Distribution}
\label{tab:ZIBNB_trigammafree_DebTrivedi_ofp_twodistances}
\end{table}

From Table~\ref{tab:ZIBNB_trigammafree_DebTrivedi_ofp_twodistances}, we can see a decreasing trend for our trigamma-free results in both the Frobenius distance and the maximum length change of confidence intervals as $M$ increases. Both distances reach zero at $M=11000$. 
As for the Monte Carlo estimator, both the Frobenius distance and the maximum length change do not seem to change much even when $M$ reaches as large as $10^{6}$. 
The difficulty of this special case actually comes from a tiny eigenvalue $1.76\times10^{-17}$ of the exact Fisher information matrix ${\mathbf G}$, which leads to a large deviation of the estimated inverse Fisher information matrix from the exact one. For our trigamma-free method, an $M$ as large as $11000$ yields an upper bound of error $P(Y > M+1)/(\hat\nu + M) = 1.94 \times 10^{-17}$ with $Y\sim {\rm BNB}(\hat\nu, \hat\alpha, \hat\beta)$ (see Theorem~\ref{thm:Psi_approximation_general}) and the exact value of error is even smaller. Actually, the calculated key components $E\Psi_1(\hat\nu + Y)$, $E\Psi_1(\hat\beta + Y)$, and $E\Psi_1(\hat\nu + \hat\alpha + \hat\beta + Y)$ with $Y\sim {\rm BNB}(\hat\nu, \hat\alpha, \hat\beta)$ stay exactly the same after $M\geq 11,000$, which makes the calculated inverse Fisher information matrix unchanged as long as $M\geq 11,000$. On the other hand, the Monte Carlo estimator may need an $M$ as large as $10^8\sim 10^9$ to achieve roughly the same accuracy level, which is seldom used in practice.

In Table~\ref{tab:ZIBNB_trigammafree_vs_MonteCarlo_DebTrivedi_ofp_timecost}, we list the computational time needed for both methods. Using our trigamma-free method, we only need $M=11000$ and $0.02$ second to achieve a satisfactory result for this case, while $M=10^6$ and $1.59$ seconds are still not enough if we use the Monte Carlo estimator.
}\hfill{$\Box$}
\end{example}

\begin{table}[ht]
\centering
\begin{tabular}{|c|c|c|c|}
\hline
\textbf{\begin{tabular}[c]{@{}c@{}}Trigamma-free\\ M\end{tabular}} & \textbf{\begin{tabular}[c]{@{}c@{}}Computation\\Time (seconds)\end{tabular}} & \textbf{\begin{tabular}[c]{@{}c@{}}Monte Carlo\\M\end{tabular}} & \textbf{\begin{tabular}[c]{@{}c@{}}Computation\\Time (seconds)\end{tabular}} \\ \hline
5000   & 0.02                                 & 5000 & 0.14 \\ \hline
10000  & 0.03 & 10000 & 0.17 \\ \hline        11000 & 0.02 & $10^5$ & 0.37  \\ \hline
12000 & 0.03 & $10^6$ & 1.59 \\ \hline        \end{tabular}
\caption{Computational Time for  
{\tt ofp} in {\tt DebTrivedi} under ZIBNB Distribution}
\label{tab:ZIBNB_trigammafree_vs_MonteCarlo_DebTrivedi_ofp_timecost}
\end{table}

\subsection{Applications to zero-inflated regression models}\label{sec:application_ZI_regression_model}

In this section, we apply our trigamma-free method for calculating the Fisher information matrices involved in zero-inflated regression models (ZIRM) and hurdle regression models (HRM).

Following \cite{hanithesis}, suppose we have independent observations $(Y_i, {\mathbf x}_i)$, $i=1, \ldots, n$ with covariates ${\mathbf x}_i = (x_{i1},$ $ \ldots,$ $x_{id})^T \in \mathbb{R}^d$ and response $Y_i \in \mathbb{R}$. 
Suppose $Y_i$ follows a zero-altered model (also called a hurdle model)
\[
f_{\rm ZA}(y| \phi_i, {\boldsymbol\theta}_i) = \phi_i\cdot  {\mathbf 1}_{\{y=0\}} + \frac{1-\phi_i}{1-p_0(\boldsymbol\theta_i)} f_{\boldsymbol\theta_i}(y) \cdot {\mathbf 1}_{\{y\neq 0\}}
\]
or a zero-inflated model 
\[
f_{\rm ZI}(y| \phi_i, {\boldsymbol\theta}_i) = \left[\phi_i + (1-\phi_i) p_0({\boldsymbol\theta}_i)\right]\cdot  {\mathbf 1}_{\{y=0\}} + (1-\phi_i) f_{{\boldsymbol\theta}_i}(y) \cdot {\mathbf 1}_{\{y\neq 0\}}
\]
where $\phi_i \in (0,1)$, ${\boldsymbol\theta}_i = (\theta_{i1}, \ldots, \theta_{ib})^T \in \mathbb{R}^b$, and $f_{\boldsymbol\theta_i}(y)$ is a baseline distribution function. In \cite{hanithesis}, both HRM and ZIRM assumes the existence of link functions $g$ and $h_1,$ $\ldots,$ $h_b$ such that
\begin{equation}\label{eq:ZARM}
\left\{\begin{array}{ll}
g(\phi_i) = {\mathbf G}_i^T \boldsymbol\gamma\ , & i=1, \ldots, n\\
h_j(\theta_{ij}) = {\mathbf B}_{ij}^T {\boldsymbol\beta}_j\ , & i=1, \ldots, n;\ j=1, \ldots, b
\end{array}
\right.
\end{equation}
where $\boldsymbol\gamma, {\boldsymbol\beta}_1, \ldots, {\boldsymbol\beta}_b$ are regression coefficients, ${\mathbf G}_i = (r_1({\mathbf x}_i), \ldots, r_s({\mathbf x}_i))^T \in \mathbb{R}^s$  and ${\mathbf B}_{ij} = (q_{j1}({\mathbf x}_i),$ $ \ldots, $ $q_{jt_j}({\mathbf x}_i))^T \in \mathbb{R}^{t_j}$  are the corresponding predictors, and $r_i$'s and $q_{ji}$'s are known predictor functions. For example, ${\mathbf G}_i = {\mathbf B}_{ij} = (1, x_{i1}, \ldots, x_{id})^T$ for main-effects models.

As described in Section~3.4 of \cite{hanithesis}, the calculation of the corresponding Fisher information matrix for both HRM and ZIRM depends on calculating 
\[
E\left[ \frac{\partial \log f_{\boldsymbol\theta_i}(Y_i')}{\partial \theta_{is}} \cdot \frac{\partial \log f_{\boldsymbol\theta_i}(Y_i')}{\partial \theta_{it}} \right]
\]
for each $i=1, \ldots, n$, and $s,t=1, \ldots, b$, where $Y_i'\sim f_{\boldsymbol\theta_i}(y)$. 

In this paper, however, we need the following equality under regularity conditions (see, for example, Section~2.5 in \cite{lehmann1998theory})
\[
E\left[ \frac{\partial \log f_{\boldsymbol\theta_i}(Y_i')}{\partial \theta_{is}} \cdot \frac{\partial \log f_{\boldsymbol\theta_i}(Y_i')}{\partial \theta_{it}} \right]
= -
E\left[ \frac{\partial^2 \log f_{\boldsymbol\theta_i}(Y_i')}{\partial \theta_{is}\partial \theta_{it}}\right]
\]
so that our trigamma-free method can be used for calculating the Fisher information matrix, when the baseline distribution is binomial, beta binomial, negative binomial, or beta negative binomial (see also  Section~\ref{sec:calculate_Fisher_information}).

\begin{example}\label{ex:ZIBNB} 
{\bf Zero-inflated beta negative binomial (ZIBNB) regression model}\quad
{\rm The zero-inflated beta negative binomial (ZIBNB) model has been introduced in \cite{aldirawi2022modeling} as a flexible probabilistic model for microbiome data study. The ZIBNB regression model is a special case of the ZIRM~\eqref{eq:ZARM}, whose baseline distribution is the beta-negative binomial distribution with parameters $\boldsymbol\theta_i = (\nu_i, \alpha_i, \beta_i)^T$ and the probability mass function (pmf) given by
\[
f_{{\boldsymbol\theta}_i}(y_i) = \dbinom{\nu_i+y_i-1}{y_i} \frac{{\rm Beta}(\nu_i + \alpha_i,  y_i + \beta_i)}{{\rm Beta}(\alpha_i,\beta_i)}
\]
where $y_i=0, 1, \ldots$, $i=1, \ldots, n$.

Given the data $\{(y_i, {\mathbf x}_i)\mid i=1, \ldots, n\}$, we {\it (i)} find the MLE $\hat{\boldsymbol{\gamma}}, \hat{\boldsymbol{\beta}}_1, \ldots, \hat{\boldsymbol{\beta}}_b$ for the model parameters with $b=3$ in this case; {\it (ii)} for each $i=1, \ldots, n$, obtain the MLE $\hat{\phi}_i = g^{-1}({\mathbf G}_i^T \hat{\boldsymbol{\gamma}})$, $\hat{\nu}_i = h_1^{-1}({\mathbf B}_{i1}^T \hat{\boldsymbol{\beta}}_1)$, $\hat{\alpha}_i = h_2^{-1}({\mathbf B}_{i2}^T \hat{\boldsymbol{\beta}}_2)$, $\hat{\beta}_i = h_3^{-1}({\mathbf B}_{i3}^T \hat{\boldsymbol{\beta}}_3)$; {\it (iii)} for each $i=1, \ldots, n$, calculate the Fisher information matrix $\hat{\mathbf F}_{iM}$ at $(\hat{\phi}_i, \hat{\nu}_i, \hat{\alpha}_i, \hat{\beta}_i)$ following the formulae in Example~S.2 in the Supplementary Materials of \cite{dousti2022r} using either our trigamma-free method or the Monte Carlo estimator with size $M$ (see also Example~\ref{ex:ZABNB_ZIBNB}); and {\it (iv)} calculate the estimated Fisher information matrix $\hat{\mathbf F}_M = n^{-1} \sum_{i=1}^n \hat{\mathbf F}_{iM}$ for the regression model.

We utilize the real data {\tt DebTrivedi} in the R package {\tt MixAll}~\citep{mixall} again for illustration purposes. In this example, we model the demand of medical insurance for $n=4,406$ individual's physician office visits ({\tt ofp}, the response). The covariates considered in this study are {\tt hosp} (the number of hospital stays), {\tt numchron} (the number of chronic conditions), {\tt school} (the number of years of education), {\tt gender}, the health status variables (self-perceived health status), and {\tt privins} (private insurance indicator).
Among different ZIRM main-effects models with different link functions, we adopt the ZIBNB regression model  with the probit link, which attains the smallest AIC and BIC values (see, for example, \cite{hastie2009elements} for a good review).

To evaluate the accuracy and efficiency of our trigamma-free method and the Monte Carlo estimator, we denote the calculated Fisher information matrix based on our trigamma-free method with $M=10^5$ as ${\mathbf G}$ and treat it as the exact Fisher information matrix. 
We choose $M=10^5$ instead of $10^6$ here because it costs less computational time and the corresponding approximations  in this example are numerically the same as the ones with $M=10^6$.
Similar as in Example~\ref{ex:ZABNB_ZIBNB}, we calculate the Frobenius distance and the maximum length change of confidence intervals between $\hat{\mathbf F}_M^{-1}$ and ${\mathbf G^{-1}}$ and list them in Table~\ref{table:ZIBNB-l}. We can see that our trigamma-free method obtains essentially the same Fisher information matrix at $M=1000$, while the Monte Carlo estimator with an $M$ as large as $20,000$ is still not comparable with the trigamma-free result with $M=100$. 

We also record the computational time of both methods in Table~\ref{table: ZIBNB-time}. It takes only 2.70 seconds for the trigamma-free method with $M=1000$, whose result is satisfactory, while it costs the Monte Carlo estimator $123$ seconds with $M=20,000$.

\begin{table}[ht]
\centering
\begin{tabular}{|c|c|c|c|}
\hline
{\bf Method} & $M$ & \textbf{Frobenius Norm} & \textbf{Max Length Change} \\ \hline
 & 100  & $0.0083$               & $0.00058$     \\ 
Trigamma-free 
 & 1000 & $0$               & $0$                         \\ 
 & 5000 & 0                       & 0                                \\  
 & 20000 & 0                       & 0                                \\ \hline
 & 1000  & $0.0609$               & $0.00899$                          \\ 
Monte Carlo
 & 5000 & $0.0352$               & $0.00604$                          \\ 
 & 20000  & $0.0289$               & $0.00494$                            \\ \hline 
\end{tabular}
\caption{Distance between the Estimated and Exact Inverse Fisher Information Matrices for {\tt DebTrivedi} under the ZIBNB Regression Model
}
\label{table:ZIBNB-l}
\end{table}

\begin{table}[ht]
\centering
\begin{tabular}{|c|c|c|c|}
\hline
\textbf{\begin{tabular}[c]{@{}c@{}}Trigamma-free\\ M\end{tabular}} & \textbf{\begin{tabular}[c]{@{}c@{}}Computation\\Time (seconds)\end{tabular}} & \textbf{\begin{tabular}[c]{@{}c@{}}Monte Carlo\\M\end{tabular}} & \textbf{\begin{tabular}[c]{@{}c@{}}Computation\\Time (seconds)\end{tabular}} \\ \hline
100   & 0.65     & 1000 & 6.11 \\ \hline
1000  & 2.70     & 5000 & 30.20 \\ \hline   
10000  & 25.20    & 20000 & 123.00 \\ \hline   
\end{tabular}
\caption{Computational Time for the Fisher Information Matrix under the ZIBNB Regression Model
}
\label{table: ZIBNB-time}
\end{table}

To further illustrate the advantage of our trigamma-free method over the Monte Carlo estimator, we summarize in Table~\ref{table: ZIBNB-CI} the confidence intervals of a particular regression parameter $\beta_{33}$ based on different methods and different $M$'s. Since the estimate $\hat{\beta}_{33} = 0.0711$, which represents the effect of {\tt school} on distribution parameter $\beta_i$, is fairly close to $0$, it is important to test whether it is significantly away from zero, which can be concluded based on a $95\%$ confidence interval.
From Table~\ref{table: ZIBNB-CI}, we can see that all the three confidence intervals based on the trigamma-free method do not contain zero, which consistently imply that $\hat{\beta}_{33}$ is significantly different from zero.
Nevertheless, the three confidence intervals based on the Monte Carlo estimator are not consistent. The interval with $M=1000$, which is commonly used in practice, contains zero, while the other two intervals do not, which implies that the test based on the Monte Carlo estimator can be erroneous due to its low accuracy even if the lengths of confidence intervals do not seem to change much as $M$ increases. 
To achieve a more reliable result when applying the Monte Carlo estimator, we need a much bigger $M$.
That being said, our trigamma-free method is not only much faster than the Monte Carlo estimator, but also more accurate and more reliable.

\begin{table}[ht]
\centering
\begin{tabular}{|c|c|c|c|c|}
\hline
{\bf Method} & $M$ & \textbf{Lower Bound} & \textbf{Upper Bound} & \textbf{$|\frac{\bf Estimate}{\bf Length}|$} \\ \hline
 & 100   & 0.00047         & 0.14176            & 0.503         \\ 
Trigamma-free & 1000  & 0.00043         & 0.14180            & 0.503         \\ 
 & 5000 & 0.00043         & 0.14180            & 0.503         \\
 & 20000 & 0.00043         & 0.14180            & 0.503         \\ \hline
 & 1000  & -0.00010        & 0.14233            & 0.499         \\ 
Monte Carlo    & 5000  & 0.00027         & 0.14196            & 0.502          \\ 
 & 20000 & 0.00039         & 0.14183            & 0.503         \\ \hline
\end{tabular}
\caption{Confidence Intervals for the {\tt school} Coefficient $\beta_{33}$ with $\hat{\beta}_{33} = 0.0711$}
\label{table: ZIBNB-CI}
\end{table}

}
\end{example}

\section{Conclusion}\label{sec:conclusion}

In this paper, we propose a highly accurate and efficient way of computing $E\Psi_1(\nu+Y)$ with a non-negative integer-valued random variable $Y$. It only needs to compute the trigamma function $\Psi_1$ once at $\nu$, which minimizes the time cost and numerical errors due to the calculation of trigamma function values. Its theoretical upper bound $P(Y>M+1)/(\nu+M)$ of the error term vanishes quickly with $M$, which makes the approximation approach the exact value quickly.

In Section~\ref{sec:improved_approximation}, we develop a calibrated approximation for $E\Psi_1(\nu+Y)$. By removing a portion $\rho_*$ of the theoretical upper bound $P(Y>M+1)/(\nu+M)$, the modified approximation $\Psi_e(\nu, M, \rho_*)$ could be more accurate than $\Psi_e(\nu, M, 0)$ when $M < E(Y)/2$. One may further reduce the approximation error for moderate $M$ by continuing to add a calibration term, which is out of the scope of this paper. 

In Section~\ref{sec:applications}, we show by examples that a highly accurate computation on $E\Psi_1(\nu+Y)$ is critical under some circumstances when the commonly used Monte Carlo estimator is not satisfactory. Overall, we recommend our trigamma-free method $\Psi_e(\nu, M, 0)$ with $M>2E(Y)$ for computing $E\Psi_1(\nu+Y)$, including but not restricted to negative binomial and beta negative binomial distributions and regression models.  

In Theorem~\ref{thm:E_Psi_1} and Corollary~\ref{col:E_Psi_1_n}, we also provide explicit formulae for $E\Psi(\nu + Y)$ with $Y$ taking non-negative integer values only. The quantity $\Psi(\nu) + \sum_{y=0}^M P(Y> y)/(\nu+y)$ can be used to approximate $E\Psi(\nu+Y)$, which is similar in spirit to $\Psi_e(\nu, M, 0)$ for approximating $E\Psi_1(\nu + Y)$.




\section*{Acknowledgments}

This research was partially supported by the U.S.~NSF grant DMS-1924859.

\section*{Statements and Declarations}

{\bf Conflict of interest}  The authors declare that they have no conflict of
interest.



\section*{Appendix}

\noindent
{\bf Proof of Theorem~\ref{thm:E_Psi_1}}\label{secA1}

According to \eqref{eq:Psi_x+y}, 
\begin{eqnarray*}
\Psi(\nu) - E\Psi(\nu +Y) &=& \sum_{y=1}^\infty P(Y=y) \left[\Psi(\nu) - \Psi(\nu+y)\right]\\
&=& -\sum_{y=1}^\infty P(Y=y) \sum_{j=0}^{y-1} \frac{1}{\nu+j}\\
&=& -\sum_{j=0}^\infty \frac{1}{\nu+j} \sum_{y=j+1}^\infty P(Y=y)\\
&=& -\sum_{j=0}^\infty \frac{P(Y> j)}{\nu+j}\\
&=& -\sum_{y=0}^\infty \frac{P(Y> y)}{\nu + y}
\end{eqnarray*}
Similarly, according to \eqref{eq:Psi_x+y_Psi_x}, 
\begin{eqnarray*}
\Psi_1(\nu) - E\Psi_1(\nu +Y) &=& \sum_{y=1}^\infty P(Y=y) \left[\Psi_1(\nu) - \Psi_1(\nu+y)\right]\\
&=& \sum_{y=1}^\infty P(Y=y) \sum_{j=0}^{y-1} \frac{1}{(\nu+j)^2}\\
&=& \sum_{j=0}^\infty \frac{1}{(\nu+j)^2} \sum_{y=j+1}^\infty P(Y=y)\\
&=& \sum_{j=0}^\infty \frac{P(Y> j)}{(\nu+j)^2}\\
&=& \sum_{y=0}^\infty \frac{P(Y> y)}{(\nu + y)^2}
\end{eqnarray*}
\hfill{$\Box$}

\noindent
{\bf Proof of Theorem~\ref{thm:Psi_approximation_general}}\label{secA2}

According to  Theorem~\ref{thm:E_Psi_1}, 
\begin{eqnarray*}  & & \left|E\Psi_1(\nu +Y) - \Psi_1(\nu) + \sum_{y=0}^M \frac{P(Y > y)}{(\nu + y)^2}\right|\\ 
&=& \left|\sum_{y=M+1}^\infty \frac{P(Y > y)}{(\nu + y)^2}\right|\\
&\leq & P(Y>M+1) \cdot \sum_{y=M+1}^\infty \frac{1}{(\nu+y)^2}\\ 
&\leq & P(Y>M+1)\sum_{y=M+1}^\infty  \left(\frac{1}{\nu+y-1} - \frac{1}{\nu+y}\right)\\
&=& \frac{P(Y>M+1)}{\nu+M}
\end{eqnarray*}
\hfill{$\Box$}

\noindent
{\bf Proof of Lemma~\ref{lem:lower_bound_difference}}\label{secA3}

First of all,
\begin{eqnarray*}
\sum_{y=M+1}^\infty \frac{P(Y>y)}{(\nu+y)^2} &=& \sum_{y=M+1}^\infty \sum_{k=y+1}^\infty \frac{P(Y=k)}{(\nu+y)^2}\\
&=& \sum_{k=M+2}^\infty \sum_{y=M+1}^{k-1} \frac{P(Y=k)}{(\nu+y)^2}\\
&=& \sum_{k=M+2}^\infty P(Y=k) \cdot \sum_{y=M+1}^{k-1} \frac{1}{(\nu+y)^2}
\end{eqnarray*}
Since
\begin{eqnarray*}
\sum_{y=M+1}^{k-1} \frac{1}{(\nu+y)^2} &<& \sum_{y=M+1}^{k-1} \left(\frac{1}{\nu+y-1} - \frac{1}{\nu+y}\right) = \frac{1}{\nu+M}-\frac{1}{\nu+k-1}\\
\sum_{y=M+1}^{k-1} \frac{1}{(\nu+y)^2} &>&  \sum_{y=M+1}^{k-1} \left(\frac{1}{\nu+y} - \frac{1}{\nu+y+1}\right) = \frac{1}{\nu+M+1}-\frac{1}{\nu+k}  
\end{eqnarray*}
The conclusion follows.
\hfill{$\Box$}

\medskip
\noindent
{\bf Proof of Theorem~\ref{thm:rho=1/2}}\label{secA4}

By the definition of $\Psi_e(\nu, M, \rho)$ and Lemma~\ref{lem:lower_bound_difference}, 
\begin{eqnarray*}
\Psi_e (\nu, M, \rho) - E\Psi_1(\nu+Y) &=& \sum_{y=M+1}^\infty \frac{P(Y > y)}{(\nu+y)^2} - \rho \cdot \frac{P(Y>M+1)}{\nu+M}\\
&\leq & (1-\rho) \cdot \frac{P(Y>M+1)}{\nu+M} 
\end{eqnarray*}
On the other hand,
\begin{eqnarray*}
& & \Psi_e (\nu, M, \rho) - E\Psi_1(\nu+Y) \\
&\geq & \frac{P(Y>M+1)}{\nu+M+1} - \sum_{k=M+2}^\infty \frac{P(Y=k)}{\nu+k} - \rho \cdot \frac{P(Y>M+1)}{\nu+M} \\
&\geq & \frac{P(Y>M+1)}{\nu+M+1} - \frac{P(Y>M+1)}{\nu+M+2} - \rho \cdot \frac{P(Y>M+1)}{\nu+M}\\
&=& \frac{P(Y>M+1)}{\nu+M}\left[ \frac{\nu+M}{(\nu+M+1)(\nu+M+2)} - \rho\right]
\end{eqnarray*}
Therefore, $|\Psi_e(\nu, M, \rho) - E\Psi_1(\nu+Y)| \leq U(\rho)\cdot \frac{P(Y>M+1)}{\nu+M}$, where 
\[
U(\rho) =  \max\left\{|1-\rho|, \left|\rho - \frac{\nu+M}{(\nu+M+1)(\nu+M+2)}\right|\right\}
\]
It can be verified that $U(\rho)$ attains its minimum when 
\[
1-\rho = \rho - \frac{\nu+M}{(\nu+M+1)(\nu+M+2)}
\]
or equivalently, $\rho = \rho_*$.
\hfill{$\Box$}


\end{document}